\documentclass[sigconf,authorversion,nonacm]{acmart}
\AtBeginDocument{%
  }

\copyrightyear{2026}
\acmYear{2026}
\setcopyright{cc}
\setcctype{by}
\acmConference[CHI '26]{Proceedings of the 2026 CHI Conference on Human Factors in Computing Systems}{April 13--17, 2026}{Barcelona, Spain}
\acmBooktitle{Proceedings of the 2026 CHI Conference on Human Factors in Computing Systems (CHI '26), April 13--17, 2026, Barcelona, Spain}
\acmPrice{}
\acmDOI{10.1145/3772318.3790709}
\acmISBN{979-8-4007-2278-3/2026/04}

\usepackage[utf8]{inputenc}
\usepackage{framed,float}
\usepackage{xcolor}
\usepackage{enumitem}
\usepackage{graphicx}
\usepackage{subcaption}
\usepackage{colortbl}

\definecolor{darkblue}{HTML}{8cbcdc}
\definecolor{lightblue}{HTML}{e0f0ff}
\definecolor{storybg}{HTML}{f0f5ff}
\definecolor{storyrule}{HTML}{000080}

\newcommand{\rev}[1]{{\color{black} #1}}

\usepackage[
  skip=2pt,
  font=small,
  width=0.8\textwidth
]{caption}

\newcounter{story25}
\begin{document}


\title{Touching Emotions, Smelling Shapes: Exploring Tactile, Olfactory and Emotional Cross-sensory Correspondences in Preschool Aged Children}



\author{Tegan Roberts-Morgan}
\email{tegan. roberts. morgan@bristol. ac. uk}
\orcid{0009-0000-1404-802X}
\affiliation{
  \institution{University of Bristol}
  \city{Bristol}
  \country{UK}
}

\author{Min S. Li}
\email{min. li@bristol. ac. uk}
\orcid{0000-0003-3673-202X}
\affiliation{
  \institution{University of Bristol}
  \city{Bristol}
  \country{UK}
}

\author{Priscilla Lo}
\email{priscilla. lo@bristol. ac. uk}
\orcid{0009-0001-3571-624X}
\affiliation{
  \institution{University of Bristol}
  \city{Bristol}
  \country{UK}
}

\author{
Zhuzhi Fan}
\email{zhuzhi. fan@bristol. ac. uk}
\orcid{0000-0002-0732-7642}
\affiliation{
  \institution{University of Bristol}
  \city{Bristol}
  \country{UK}
}

\author{Dan Bennett}
\orcid{0000-0002-9330-5529}
\email{dan.bennett@bristol.ac.uk}
\email{dtbe@cs.aau.dk}
\affiliation{
  \institution{University of Bristol}
  \city{Bristol}
  \country{UK}
}
\affiliation{
  \institution{Aalborg University}
  \city{Aalborg}
  \country{Denmark}
}

\author{Oussama Metatla}
\email{o. metatla@bristol. ac. uk}
\orcid{0000-0002-3882-3243}
\affiliation{
  \institution{University of Bristol}
  \city{Bristol}
  \country{UK}
}

\renewcommand{\shortauthors}{Roberts-Morgan et al. }
\renewcommand{\shorttitle}{Touching Emotions, Smelling Shapes}
\begin{abstract}

The use of a wide range of sensory modalities is increasingly central to technologies for learning, communication, and affective regulation. During the preschool years, sensory integration develops rapidly, shaping how children perceive and make sense of their environments. A key component of this process is cross-sensory correspondence: the systematic ways in which perceptions in different sensory modalities influence one another. Despite its relevance, little is known about cross-sensory correspondences in preschool-aged children (2-4 years). We present a study with 26 preschoolers examining smell-touch-emotion correspondences through playful tasks. We found significant correspondences both between sensory modalities and between sensory modalities and affective judgements. Further analysis revealed association strategies underpinning these mappings. We contribute empirical insights into cross-sensory correspondences in early childhood, design guidelines that align with how preschoolers relate sensory input, and a replicable method for probing cross-sensory cognition in this age group.

\end{abstract}


 \begin{teaserfigure}
   \includegraphics[width=\textwidth]{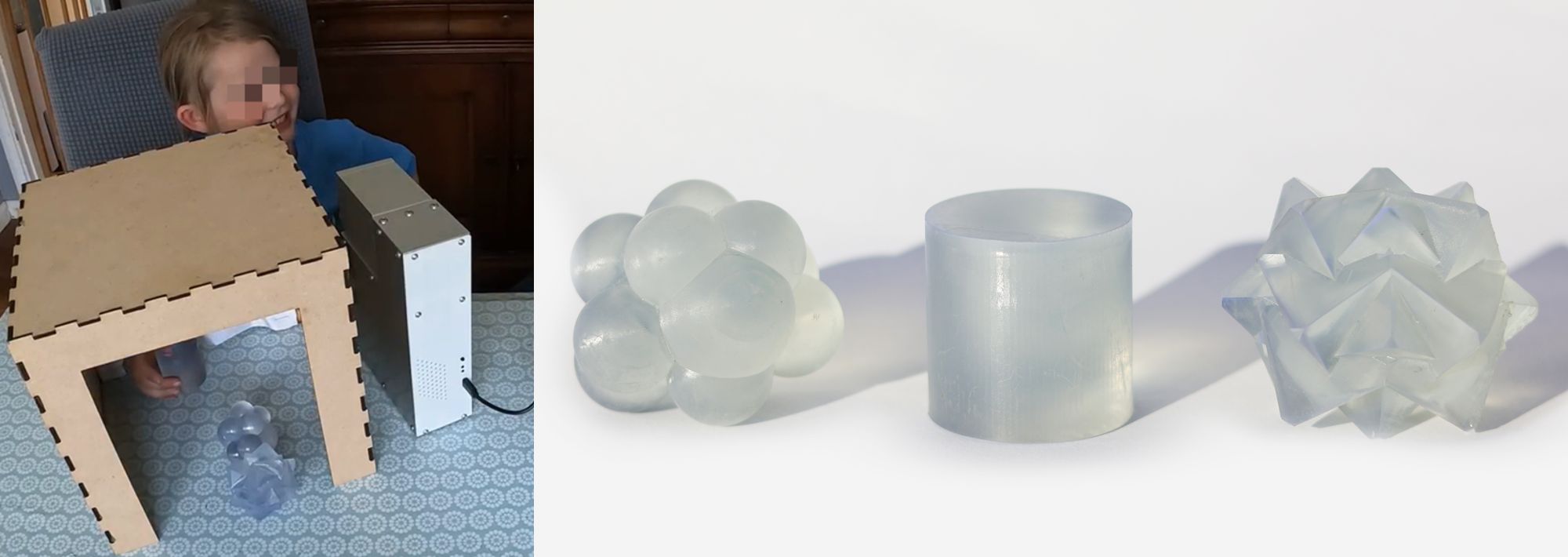}
   \caption{Left: A child participant engaging in the study with the experimental setup. Right: The three translucent tactile stimuli used in the study, \textit{Round}, \textit{Cylinder} and \textit{Spiky}.}
   \Description{The figure is divided into two parts. On the left, a child is sitting at a table, smiling, and interacting with a wooden box setup that contains the tactile stimuli. On the right, three translucent objects are shown side by side: the first looks like several smooth balls joined together, the second is a sharp-edged star-like geometric shape, and the third is a simple, smooth cylinder. These represent the tactile stimuli the children explored during the study.}
   \label{fig:teaser}
 \end{teaserfigure}

\maketitle

\section{Introduction}
Everyday experiences are inherently multisensory
\footnote{We distinguish between two related notions of \emph{multisensory} and \emph{cross-sensory} interaction. \emph{Multisensory interaction} broadly refers to systems that integrate multiple sensory modalities (e.g., vision, audition, somatosensation, olfaction, gustation) for input and output, often to enhance immersion, realism, or accessibility \cite{Obrist_Velasco_Vi_Ranasinghe_Israr_Cheok_Spence_Gopalakrishnakone_2016}. We use the phrase \emph{Cross-sensory interaction} to refer to design practices that intentionally leverage systematic correspondences between sensory modalities, moving beyond the co-presentation of multiple senses to purposefully shape perception and experience through their interaction \cite{likepopcorn, feelingcolours, steer2024squishy}. }.
We interpret and interact with the world through sight, sound, touch, taste, smell, and beyond \cite{odorop, spence2011multisensory}. Yet most digital technologies remain focused on visual, auditory, and tactile interaction, with chemical senses (smell and taste) rarely incorporated \cite{smellimportant, vision, smellunder}. This limitation is important given the potential for broader sensory engagement to enhance learning \cite{shams2008benefits}, support accessibility \cite{magika, hoomie}, and enrich children’s playful interactions \cite{theresia2021applying}. Olfaction in particular remains underexplored in HCI \cite{odorop, smellnoreview}, despite evidence of its key role in memory, affect, and social connection, and recent advances in olfactory design and display \cite{smell-message}. 

Maggioni et al. \cite{smellstruggles} highlight several key challenges in integrating chemical senses into technologies, including limited scent delivery technologies, individual differences in olfactory sensitivity, and a lack of understanding of how chemical senses interact with other modalities. One promising approach to addressing the latter challenge is to investigate cross-sensory correspondences: stable and non-arbitrary mappings between features across senses. While prior research has explored olfactory correspondences in children aged 10–17 \cite{likepopcorn}, there is evidence that different age groups differ in how they form and communicate associations between olfactory and other experiences.
\citet{bluesalt} observed that older children and adults employ a wider repertoire of cross-sensory metaphors than younger groups. While there is evidence that preschool-aged children (ages 2-4)  are sensitive to correspondences in vision and audition \cite{tastevision, OZTURK2013173, nava2016audio}, correspondences involving olfaction and emotion have been largely unexamined. This gap is significant because early childhood represents a critical window for the development of sensory integration \cite{MASON201948, dionne2015multisensory}, when the foundations of cross-sensory cognition are rapidly developing. Understanding how young children link smell, touch, and emotion can thus provide both theoretical insight into early perceptual development and practical guidance for integrating a richer spectrum of sensory modalities in technology design.  

To address this gap, we conducted a study with 26 preschool children (ages 2-4), combining playful methods with narrated tasks to probe cross-sensory correspondences. Children were presented with tactile, olfactory, and affective stimuli and asked about their associations between them. Our results indicate systematic correspondences: a \textit{spiky} 3D stimulus was most strongly associated with a lemon scent and the label ``Kiki'', while the \textit{rounded} 3D stimulus was associated with a vanilla scent and labelled ``Bouba''. Lemon was associated with excitement, and vanilla with calm, while a neutral cylinder shape was most strongly associated with air. In further analyses, we probed and analysed the 
cross-sensory association strategies that preschool aged children used to underpin their choices.  

We conclude that researchers should adopt a more systematic approach to sensory stimuli: addressing a standardised set of sensory stimuli and drawing on cross-sensory correspondences as a principled foundation for supporting emotion communication with children. Building on this perspective, we contribute: (1) empirical evidence of smell–touch–emotion correspondences in preschool children, (2) methodological insights demonstrating the value of narrated, child-centered approaches for eliciting reasoning in this age group, and (3) design implications for cross-sensory technologies that support children’s communication and emotional expression by grounding interactions in cross-sensory correspondences.

\section{Background}
\subsection{Multisensory Child-Computer Interaction}
Multisensory interaction in Human–Computer Interaction (HCI) refers to the deliberate engagement of multiple senses to enhance communication, interaction, and user experience \cite{MultisensoryHCI}. While it is common for modern technologies to involve more than one modality, research in multisensory interaction goes further by investigating how different sensory inputs are combined and interpreted \cite{cornelio2021multisensory}. Rather than simply layering modalities, the field addresses the systematic interactions between modalities, and how they shape perception and interpretation \cite{meyer2011multisensory}.

\rev{Recent design research has emphasised that this systematic, integrated multisensory approach is key to evoking our most resonant, impactful experiences, to support reflection, focus, and engagement \cite{MultisensoryHCI, SmilesSummonWarmth}. The approach shows} particular promise in design for children, where such sensory-rich environments can foster learning \cite{shams2008benefits}, sustain engagement \cite{theresia2021applying}, and support developmental processes \cite{dionne2015multisensory}. Many toys, educational tools, and interactive environments already combine tactile, auditory, and visual cues to create immersive experiences for children \cite{Alma, multiEdu, MulisenseEnviroments, mixed, magika, hoomie}. During the preschool years, multisensory integration is still maturing: studies show that while young children already bind audiovisual speech cues, more complex correspondences continue to refine throughout early childhood \cite{DionneDostie2015MSI,karibayeva2019multisensory}. This rapid developmental window is critical, as combining inputs across senses helps children make sense of the world and supports communication, understanding, and emotional expression \cite{Cosentino2025,10040904, mastinu2023emotional}. 

Despite this promise, some sensory modalities remain underexplored within the space of child-computer interaction (CCI). In particular, olfaction has been relatively neglected in CCI research, despite smell being closely tied to early development in memory, emotion, and behaviour \cite{smellimportant}. \rev{ Recent work has clarified the potential for olfactory stimuli in interactive systems to encourage positive food attitudes \cite{puzzle}, for example, or support sleep \cite{smellnoreview}. Meanwhile, the coordination of olfactory stimuli with sight, sound or heat has been seen as key in creating emotional technologies that support reflection, awareness and communication \cite{fengItsTouchingUnderstanding2022, ThermalCards}.
Still, without a theoretical understanding of how olfaction integrates with these other modalities, there is a risk that such efforts will remain ad hoc. Addressing this gap requires a systematic understanding of how children combine and interpret sensory information across modalities. We proposed \emph{cross-sensory correspondences} can provide a strong theoretical lens for grounding the integration of a broader range of sensory modalities in child-computer interaction.}

\subsection{Cross-sensory Correspondences \& Early Development}
Cross-sensory correspondences are stable, non-arbitrary associations that link features across different sensory modalities \cite{likepopcorn}. Some correspondences arise from physical covariations in the environment, while others reflect more abstract associations, yet many are remarkably consistent across populations \cite{CrossmodalCorrespondeces}. A well-studied example is the visual-linguistic cross-sensory correspondence known as the Bouba–Kiki effect, in which 2D angular shapes are reliably associated with the word ``Kiki’’ and rounded shapes with the word ``Bouba’’ \cite{kohler1947gestalt}. 
This phenomenon is robust across cultures and languages \cite{boubakikiculture} \rev{emerging, developmentally, with language use:} preschoolers as young as 2.5 years showing the effect \cite{toddlerbouba}, while prelexical infants do not \cite{nobouba}. \rev{Further, the Bouba–Kiki effect extends beyond vision and audition and has inspired a wide range of HCI research addressing olfactory \cite{robotsmell, smellshape}, tactile \cite{steer2024squishy, likepopcorn, masie,fengItsTouchingUnderstanding2022}, colour \cite{feelingcolours}, emotion \cite{Nianmei}, and taste \cite{bubbletea, tasteBK} modalities.}

\begin{figure}[htbp]
    \centering
    \captionsetup{width=0.9\linewidth}
    \includegraphics[width=0.65\linewidth]{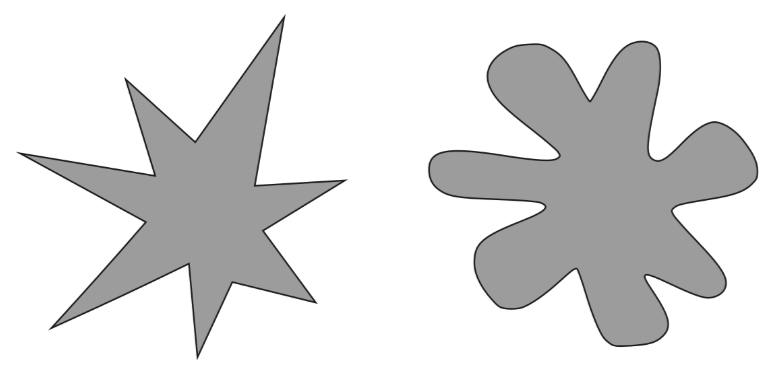}
    \caption{95\% to 98\% of the population believe that the shape on the left is associated with ``Kiki'' and the shape on the right is ``Bouba'' \cite{bk}.}
    \label{fig:boubaKiki}
    \Description{The image shows two abstract shapes side by side. The left shape is sharp and spiky, with several long, thin points sticking out in different directions. The right shape is softer and rounder, with smooth, curved edges that bulge outward.}
\end{figure}

Beyond Bouba–Kiki, prior work has documented correspondences linking touch, smell, and emotion in both adults and older children. Examples include tactile–olfactory mappings \cite{spencesmell}, tactile–emotion associations \cite{feelingcolours}, and olfactory–emotion pairings \cite{likepopcorn}. Across these studies, a small set of stimuli recur, particularly rounded versus angular shapes and scents such as lemon and vanilla. Research with younger children has been more limited, focusing mainly on vision–audio \cite{OZTURK2013173}, audio–tactile \cite{nava2016audio}, and some taste–vision correspondences \cite{tastevision}. Collectively, these findings suggest that young children are sensitive to certain cross-sensory mappings, but the scope of research remains narrow. In particular, little is known about how correspondences involving olfaction, touch, and emotion emerge during early childhood preschool years and how such correspondences could be leveraged in the design of more engaging and effective child-computer interaction.

\subsubsection{Cross-sensory Association Strategies}
Within this space, prior HCI work has progressively articulated the notion of \emph{cross-sensory association strategies} as a way to characterise how people explain, justify, and communicate the cross-sensory connections they make between stimuli from different sensory modalities. This notion was first introduced as an analytic lens to capture participants’ verbal rationales for cross-sensory associations, revealing recurring patterns in how sensory links are constructed and explained \cite{likepopcorn}. Subsequent work refined this lens by broadening and systematising these patterns, showing how such strategies operate across different sensory pairings and emotional dimensions, and how they shape interpretation in tangible interaction contexts \cite{feelingcolours}. More recent work further validated and consolidated these strategies through large-scale, generationally diverse studies, demonstrating their robustness, variability, and communicative role in the construction of cross-sensory metaphors \cite{senseonary, bluesalt}. \autoref{Descriptions} summarises the categories considered in this study, together with definitions and illustrative examples drawn from this body of work \cite{bluesalt}.
Studies with children aged 8–11 found that they most often relied on \textit{Familiar Experience}, drawing on objects or events from everyday life\rev{. In contrast, young} adults and older adults employed a wider variety of strategies \cite{bluesalt}. We adopted these categories because they capture both the content of children’s associations (e.g., naming a familiar object or event) and the expressive modes through which they communicated them (e.g., making a gesture or vocalising a sound). Importantly, these strategies are not mutually exclusive; participants can combine several at once when explaining a cross-sensory correspondence.
We extend this line of research in this paper by examining how preschool-age children exhibit association strategies with cross-sensory interaction tasks.

\begin{table*}[ht]
    \centering
    \renewcommand{\arraystretch}{1.2}
    \setlength{\tabcolsep}{5pt}
    \small 
    \begin{tabular}{|p{2.5cm}|p{6cm}|p{5.5cm}|}
        \hline
        \rowcolor{darkblue}
        \textbf{Strategy} & \textbf{Description} & \textbf{Example} \\ \hline
        
        Embodied Action & Gesturing with hand or body to help a description. & ``It feels like this,'' then stroking the floor. \\ \hline

        Grasping for Another Sense & Words from a sensory modality other than that selected for the round. & ``This tastes strong'' (e.g., referring to a shape). \\ \hline

        Personal Connection & They use a specific, personal story to describe the item. & ``This sounds like when you’re at Ikea and you get a receipt and that sound when it comes out.'' \\ \hline
        
        Sensory Features & Features of a sense have been used to describe the item. & ``Sharp and smooth.'' \\ \hline
        
        Valence & Use of terms to denote positive and/or negative qualities. & ``This tastes horrible.'' \\ \hline
        
        Vocalisation & A sound/noise is made instead of using words to describe an item. & ``This sounds like Krrrrr and tsssss.'' \\ \hline
        
        Familiar Experience & A description was created by relating the item to a common object, emotion, texture, etc. & ``This smells like a banana smoothie.'' \\ \hline
    \end{tabular}
    \caption{The seven different association strategies, their descriptions and an example. Taken from \cite{bluesalt}.}
    \Description{Table showing the seven different association strategies. These were taken from prior work by Roberts-Morgan et al. Embodied Action: Using gestures with the hand or body to support a description. For example, someone says, ``It feels like this,'' while stroking the floor. Grasping for Another Sense: Drawing on a sensory modality other than the one being tested. For instance, describing a shape by saying, ``This tastes strong.'' Personal Connection: Relating the item to a personal story or memory. An example is, ``This sounds like when you’re at Ikea and you get a receipt and that sound when it comes out.'' Sensory Features: Describing the item using features of the relevant sense, such as ``Sharp and smooth.''Valence: Using words to express positive or negative qualities, such as ``This tastes horrible.'' Vocalisation: Making a sound or noise instead of using words, for example, ``This sounds like Krrrrr and tsssss.'' Familiar Experience: Relating the item to a common object, emotion, or texture, such as ``This smells like a banana smoothie.''}
    \label{Descriptions}
\end{table*}

\subsection{Smell, Touch, and Emotion in Child-Computer Interaction}

Smell plays a central role in shaping memory, behaviour, and mood. Scents can trigger memories \cite{willander2006smell}, influence behaviour \cite{low2008scent}, and modulate affective states \cite{low2008scent}. They are also consciously used in everyday life to communicate meaning and emotion. For example, to create welcoming atmospheres \cite{ward2003ambient}, or align perfumes with particular moods or seasons \cite{mensing2023psychology}.  Recent HCI research has begun to explore olfactory stimuli as a way to enrich user experiences, including work in CCI where smell has been introduced to support engagement and learning \cite{puzzle}. Yet, olfaction remains an underdeveloped channel compared to vision, sound, or touch, especially in systems designed for preschool-aged children. 

Touch, meanwhile, is now ubiquitous in everyday technology, from touchscreens to haptic feedback. Accessibility guidelines also highlight tactile feedback as an important channel for inclusive technologies. In CCI,
tangible user interfaces (TUIs) are common, allowing children to hold, move, and manipulate physical objects \cite{paulmarshall}. TUIs are particularly suited for preschoolers 
since research emphasises the role of physical interaction in learning, for example,
supporting longer and more meaningful engagement compared to
tablet-only interactions \cite{tangiblebetter}. Beyond interaction design, there is evidence that tactile qualities such as softness, smoothness, and spikiness are associated with distinct emotions and shapes \cite{feelingcolours, steer2024squishy, etzi2016sandpaper,fengItsTouchingUnderstanding2022}. In multisensory systems, touch often provides an anchor upon which visual, auditory, or olfactory cues are layered \cite{magika, hoomie}.

Finally, emotions are central to communication and development, yet children, particularly preschoolers, often find emotions difficult to articulate verbally, as language skills are still emerging \cite{Grosse2021}. Research has shown that children frequently struggle to identify or categorise complex emotions, and that this can pose challenges for child–adult communication \cite{KestenbaumGelman1995, bender2011young, Knothe2023}. To address this, a range of technologies has been developed to scaffold children’s emotional expression. Examples include ChaCha, a chatbot that encourages children to share their feelings \cite{chacha}; MoodGems, a set of displays that allow children to record and communicate their well-being with parents \cite{moodgem}; and EmotionBlock, a tangible toolkit of wooden blocks that supports emotional management and storytelling \cite{EmotionBlock}. These systems demonstrate the importance of designing developmentally appropriate tools that help children externalise and communicate their emotions, often by relying on multisensory and embodied forms of interaction.

Despite this body of evidence for the role of touch, smell and emotion in interactive technologies, we currently lack an understanding of how preschoolers form associations between tactile and olfactory features, or connect them to emotions.
This leaves it unclear how to develop meaningful, engaging, emotionally expressive technologies for children of these ages.





\subsection{Experimental Methods from Cross-sensory Research with Preschoolers}
Most research on cross-sensory correspondences has focused on older age groups, typically between 10 and 40 years. These studies often ask participants to evaluate a range of stimuli and rate or rank how strongly they associate them across modalities \cite{spencesmell, likepopcorn, feelingcolours}. While effective for adults, such methods are not suitable for preschoolers, as they demand abstract reasoning and sustained attention, both of which are still developing between ages 2-4 \cite{gathercole2008working, Halford2007Capacity, piaget1952origins}.

In HCI, researchers have explored methods tailored to preschoolers, balancing robust data collection with sensitivity to their cognitive and emotional needs. Observational approaches are common, often situated in natural contexts such as homes, classrooms, or play spaces to capture authentic behaviours \cite{smartglasses}. Participatory design 
work for this age group has made use of tangible props \cite{tangiblepreschool} and storytelling activities \cite{storypreer}. Meanwhile, controlled experiments have typically focused on short and game-like tasks to sustain engagement \cite{preschoolrobot, yip}. To support children’s comfort and interpretation, parents or teachers are often involved as co-participants, providing reassurance and scaffolding \cite{teachandparents}. 

In studies of emotion in particular, approaches such as the Self-Assessment Manikin (SAM) \cite{SAM} and the circumplex model of affect \cite{russell1980circumplex} have been common in studies of older people. These measure affect via continuous dimensions of valence (pleasant–unpleasant), arousal (calm–excited), and dominance.

However, for younger children, the circumplex can be abstract and hard to understand.
\rev{Efforts have been made to develop more accessible forms of SAM for younger children and children with disabilities \cite{SAMAdapt}, but formal validation of such measurement instruments with school-aged children is difficult. To date
there is relatively little evidence validating the use of SAM and its variants for pre-school-aged children.
Prior work with younger children has tended to employ} simpler approaches such as binary choice \cite{likepopcorn} (e.g., happy vs. sad, excited vs. calm, in control vs. out of control) or non-verbal matching tasks \cite{matching, deak2002matching}. These approaches typically reduce the number of options to avoid overwhelming young children \cite{schupak2019choice, stockall2012right}. 

%
Building on the above literature, our study combines multiple strategies into a method that is developmentally appropriate yet capable of capturing meaningful cross-sensory data. We introduce abstract concepts through storytelling, use a teacher as a co-participant to provide reassurance and scaffolding, and adapt the SAM framework into a simpler \rev{binary choice format. We chose this approach because we felt it was easier to incorporate into an interactive storytelling format. This approach removed the focus on multiple discrete options and instead let us encourage children to express themselves in their own words and explain the reasoning behind their choices.} These design choices respect the cognitive, attentional, and communicative capacities of 2-4 year olds while still retaining standardisation, control, and comparability to results in prior research. 

\section{Study}
Our study addresses two research questions. First: what are the cross-sensory correspondences that children aged 2–4 form between smell, touch, and emotion? Second: What strategies do children aged 2-4 use to create connections between smell, touch, and emotion? We use structured, story-based conversations to explore how children, aged 2-4, associate tactile stimuli with names, smells, and emotions, and how they link specific scents to emotions. The study received ethical approval from the author's institution's ethics board. 

\subsection{Participants}
A total of 26 children (16 females, 10 males) between the ages of 2 years and 4.5 years old (M=45.08 \rev{months}, SD=9.19 \rev{months}) participated in this study, \rev{all of whom were fluent in English}. 
Participants were recruited by asking four different nurseries to send forms and information sheets to caregivers, capturing consent and demographic information. 
Caregivers were informed that essential oils would be used during the study and what their child would be taking part in, and consented to video and audio recording of the sessions. We also obtained each participant's assent for participation and recording before the study began.

\subsection{Apparatus} 
\begin{figure}[htbp]
    \centering
    \includegraphics[width=\linewidth]{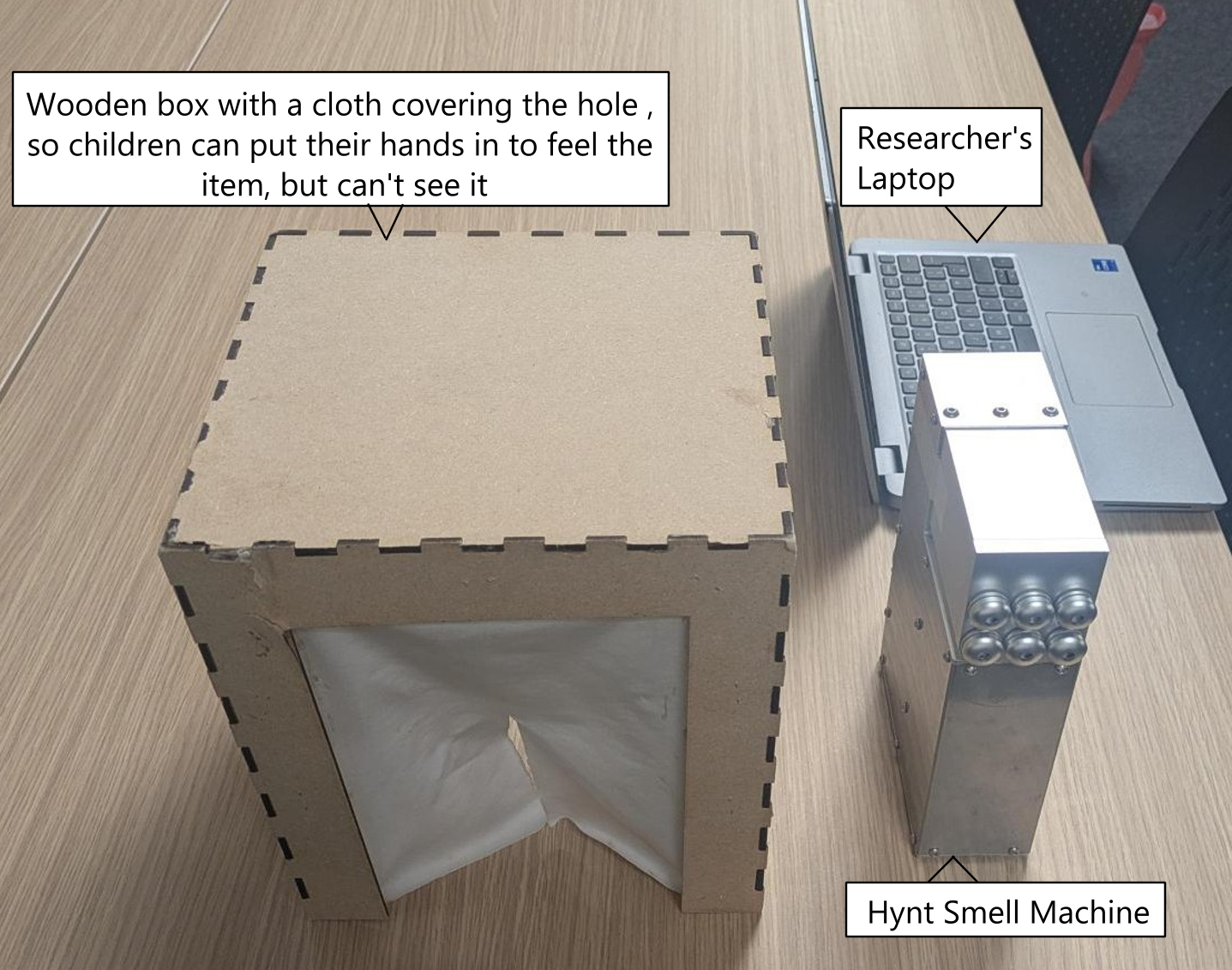}
    \captionsetup{width=0.9\linewidth}
    \caption{The child's view during the study. There is a wooden box, a scent-dispensing machine and finally the researcher's laptop. There is cloth covering the gap at the front of the wooden box, so the child can't see the tactile stimuli. The tactile stimuli are behind the researcher's laptop and can be moved easily into the box.}
    \label{fig:setup2}
    \Description{A laptop is placed on a desk with its screen open, facing the researcher’s chair. To the left of the laptop, there is a rectangular metal box connected with a cable, and behind it, a wooden box with the back and front open. The child sees the front of the wooden box, which has an opening covered by white fabric curtains that can be pushed aside to reach inside. }
\end{figure}

\begin{figure}[htbp]
    \centering
    \includegraphics[width=0.8\linewidth]{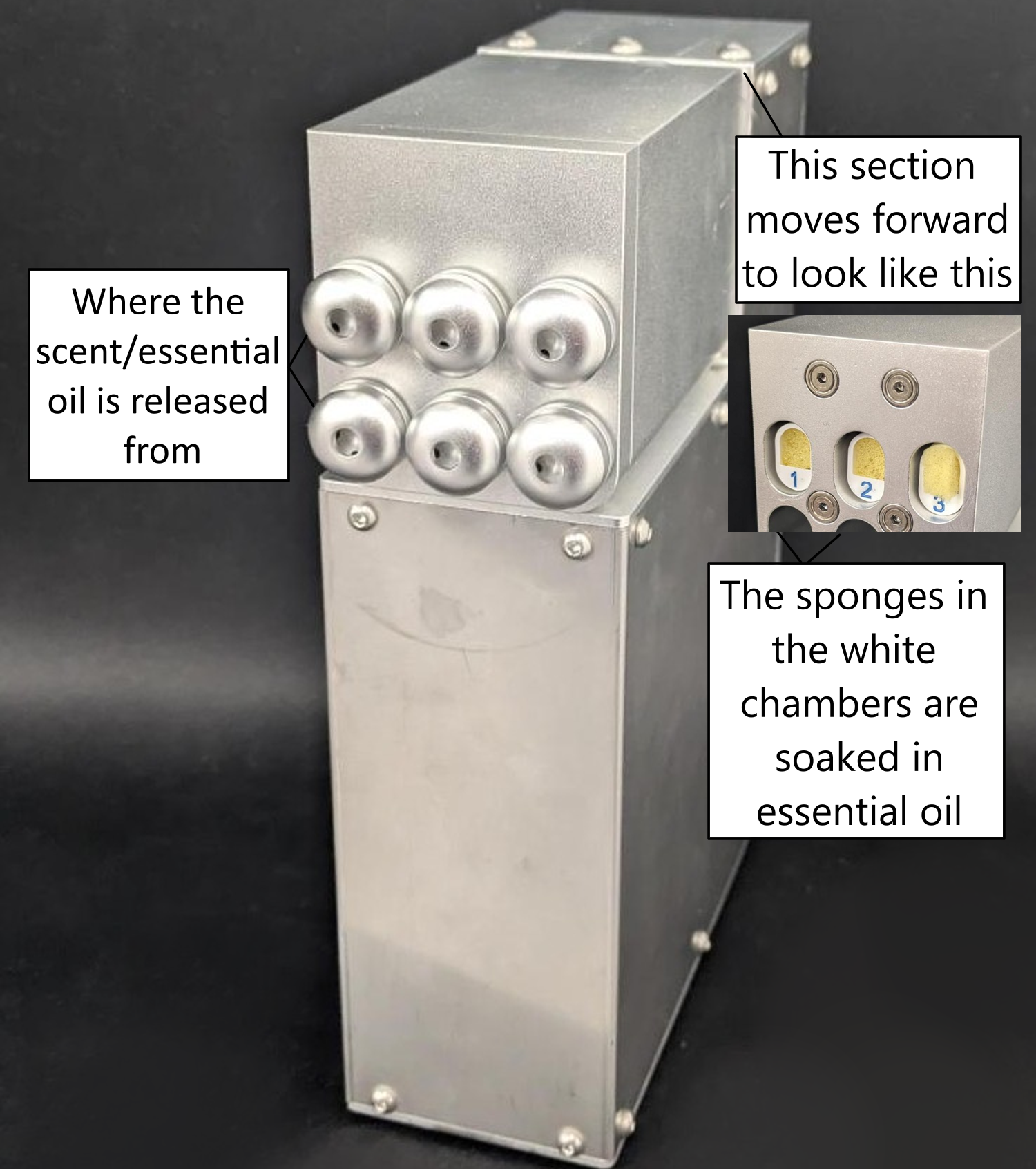}
    \captionsetup{width=0.9\linewidth}
    \caption{The HyntSmell Machine, with labels showing where the scent is dispensed from.}
    \label{fig:Hynt}
    \Description{The Hynt Smell machine, with labels saying how the scent is dispensed.}
\end{figure}

The setup used for the study is pictured in Figure \ref{fig:setup2} and \autoref{fig:Hynt}. We presented tactile stimuli to the children using a specially designed wooden box. The box was closed, with two holes: one for the experimenter to insert stimuli, and one for the child to reach in and explore the stimuli. The latter was covered in cloth to prevent children from seeing the tactile stimuli. 
Before being presented with the stimuli, the scents were dispensed using a Hynt scent-dispensing machine \cite{HyntTechnology2025}, shown in Figure \ref{fig:Hynt}, placed next to the box and controlled by a researcher using a laptop. 
The machine delivers scents from sponges, which can be soaked with essential oils or left neutral. Each child was exposed to each scent for 4 seconds; and given the option to resmell the scent at any point during the study. 
Since it is known that olfactory sensing declines in response to multiple stimuli, breaks were built into the experiment between delivering scents. 

\subsection{Stimuli}
\begin{figure}
    \centering
    \captionsetup{width=0.9\linewidth}
    \includegraphics[width=\linewidth]{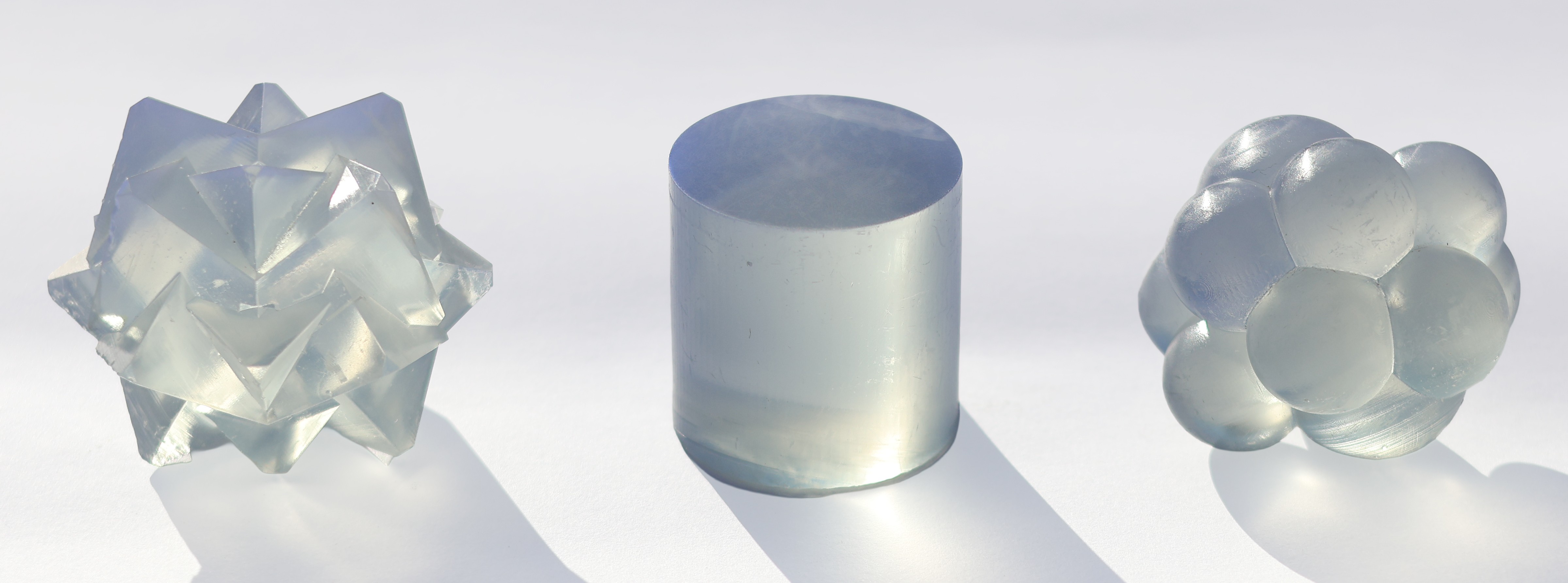}
    \caption{Tactile Stimuli. From left to right are Spiky, Cylinder and Round. }
    \Description{There are three clear, 3D, smooth objects next to each other. The first, on the left, is the Round stimulus, which is made up of several smooth spheres clustered together. The middle is the Spiky stimulus, which has sharp jagged triangles sticking out in all directions. Finally, on the right is the Cylinder stimulus, which is a simple upright cylinder shape.}
    \label{fig:tactile}
\end{figure}

For each category of stimuli (scent, touch and emotion) we used 3 different stimuli, aiming not to overwhelm the children with choice. These stimuli were selected based on prior research \cite{likepopcorn} related to the bouba-kiki phenomenon and established cross-sensory correspondences. For the olfactory stimuli, we selected lemon, vanilla, and air, as previous research has identified that lemon and vanilla are associated with angular and round shapes, respectively. 
The emotional categories were `excited', `calm', and `neither'. Prior research in children aged 11-14 has found that excitement and calm are associated with angular and rounded objects and lemon and vanilla scents, respectively. For the tactile stimuli, we chose a spiky shape, a round shape and a cylinder. A previous study in children aged 10-17 found olfactory and affective associations with these 3D shapes. 
Shapes were 3D printed using a Form4 high precision Stereolithography (SLA) 3D printer~\cite{Form4}, transparent resin~\cite{Resin} and the highest print resolution of 0.025 mm, ensuring smooth, high-quality surfaces.
Building on prior studies ~\cite{likepopcorn, bluesalt}, we used three different shape stimuli --- i.e, \textit{Round}, \textit{Spiky} and \textit{Cylinder} (see \autoref{fig:tactile}). Since we wished the stimuli to be picked up and explored, we did not use flat-bottomed shapes as used in previous studies ~\cite{bluesalt,10.1145/3544548.3580830} 
since this had the potential to present confounding haptic cues when explored from some angles.
We sized our stimuli to be easily held and manipulated by children with their whole hand: Round[54.7 mm $\times$ 54 mm $\times$ 55 mm], Spiky[56.7 mm $\times$ 57.2 mm $\times$ 66 mm] and Cylinder [43.7 mm $\times$ 43.7 mm $\times$ 43.7 mm]. A have approximately the same volume (65 mL) and the same weight (80g). To avoid discomfort or risk, sharp edges were sanded smooth. To facilitate reproduction, 3D models are provided in supplementary materials. 

\subsection{Procedure}

\begin{figure*}[htbp]
\centering

\refstepcounter{story25}
\label{story25}

\setlength{\fboxsep}{0pt} 

\fcolorbox{storyrule}{storybg}{%
\begin{minipage}{0.8\textwidth}

\colorbox{storyrule}{%
  \begin{minipage}{\linewidth}
  \setlength{\fboxsep}{16pt} 
  \setlength{\leftskip}{10pt}
  \vspace{6pt}              
  \color{white}\bfseries
  Story for Participants aged 2--4
  \vspace{6pt}              
  \end{minipage}
}

\vspace{4pt}

\begin{minipage}{\linewidth}
\setlength{\parskip}{4pt}
\setlength{\leftskip}{10pt}
\setlength{\rightskip}{10pt}

\small

\textbf{Section One - Name Each Tactile Stimuli} \\
Today, we have three characters who want to meet you! They can be called ``Kiki'', ``Bouba'', or neither of those and you can create a name! To meet these special characters, you’ll need to put your hands in this magic box and feel each one; you can't look inside the magic box. You can close your eyes if you like. (Repeated for \textit{Spiky}, \textit{Round} and \textit{Cylinder})

\begin{itemize}
    \setlength\leftskip{10pt} 
    \setlength\rightskip{10pt} 
    \item Here is this character. [Child places hand in box and is handed a tactile stimulus] Can you feel it? How does it feel?
\end{itemize}

Now, let’s play a guessing game: (Repeated for \textit{Spiky}, \textit{Round} and \textit{Cylinder})

\begin{itemize}
    \setlength\leftskip{10pt} 
    \setlength\rightskip{10pt} 
    \item What do you think this character is called [Child places hand in the box and is given a tactile stimulus to feel], ``Bouba'', ``Kiki'' or neither of those? [If the child chose ``neither'', they can then give their own name.] Why do you think that?
\end{itemize}

\noindent\dotfill

\textbf{Section Two - Tactile Stimuli and Emotion} \\
Our characters come from three magical villages. Each village is filled with a single, powerful emotion.

There’s the excited village, full of energy. The calm village is where it’s peaceful. And there’s one more village, where they don’t feel either excited or calm. Can you help me figure out which village each character is from? (Repeated for \textit{Spiky}, \textit{Round} and \textit{Cylinder})

\begin{itemize}
    \setlength\leftskip{10pt} 
    \setlength\rightskip{10pt} 
    \item Which village do you think this character is from? [Child places hand in box and is given a tactile stimulus] Excited, calm or neither? Why?
\end{itemize}

\noindent\dotfill

\textbf{Section Three - Olfactory and Emotion} \\
One day, the characters discovered a big secret about the villages they’re from: Each village only feels one type of emotion, and the emotion comes from a magical potion that fills the air!

We have three potions, each with a special scent. Can you help us match each one to the right village? (Repeated for \textit{Lemon}, \textit{Vanilla} and \textit{Air})

\begin{itemize}
    \setlength\leftskip{10pt} 
    \setlength\rightskip{10pt} 
    \item We have this one [Hynt machine disperses one of the olfactory stimuli], which village is that from? Excited, calm or neither? Why?
\end{itemize}

\noindent\dotfill

\textbf{Section Four - Tactile Stimuli and Olfactory} \\
Now our three characters want to go painting. They have lots of fun painting together, paint is everywhere! We’ve made a beautiful mess together. But we’re covered from head to toe in paint! It’s time for a bath.

We have three different soaps that our characters can choose from. Each one has a unique scent. [Hynt machine disperses one of the olfactory stimuli] Can you smell this one? What do you think this one smells like? (Repeated for \textit{Lemon}, \textit{Vanilla} and \textit{Air})

Now, we need to figure out which soaps our character uses. (Repeated for \textit{Spiky}, \textit{Round} and \textit{Cylinder})

\begin{itemize}
    \setlength\leftskip{10pt} 
    \setlength\rightskip{10pt} 
    \item Which soap would this character use? [Child places hand in box and is given a tactile stimulus.] Why?
\end{itemize}

\noindent\dotfill

\textbf{Section Five - Conclusion} \\
Each character scrubs and scrubs until all the paint is gone. Now they’re fresh and clean!

Our characters wave goodbye to you. And they say, \textit{``Thank you for helping us get clean and ready for more adventures!''}

\vspace{6pt}
\end{minipage}
\end{minipage}
}

\end{figure*}

During the study, each participant was told the story reproduced in the text box \textit{Story for Participants aged 2-4}
on a one-to-one basis. 
 \rev{We spent considerable effort on creating a suitable story, going through multiple iterations.
The story was designed to approximate experimental study methods with adults and not meaningfully influence the children's associations,
while remaining engaging and enjoyable.}

 \rev{Following an approach found in prior research, we chose to present the shapes as anthropomorphic characters, since children respond well to this \cite{goldman2024children}, and this also made it more intuitive to discuss the shapes' associations with other senses -- e.g. in terms of why the shape is named this way, its preferences, and the kinds of environment it might be found. 
We considered several options for representing emotions, for example, as pets that could feel excited, calm or neither, but rejected this since we expected children's experiences with their own pets (e.g. an excitable puppy or a calm older dog) to introduce bias. We ultimately chose to represent emotions as places where the shapes might live. We felt this slightly more abstracted approach would introduce less bias based on prior experience (while children will have only one or two pets, they will experience many places), while still being intuitive (people easily form emotional associations with places and situations \cite{manzoHouseHavenRevisioning2003}). To introduce scents, we imagined washing hands with soap. This is a highly familiar, naturalistic activity for children, where a variety of different scents can be encountered. We felt that the everydayness of the activity reduced the likelihood of particular, strong, emotional associations.} 

The story involved four tasks: first, they named the tactile stimuli, choosing either ``Bouba'', ``Kiki'' or ``neither'' with the option to give another name if ``Bouba'' or ``Kiki'' did not fit. They were then asked, in a randomised order, to smell three different scents and assign them to three tactile stimuli, assign three different emotions to the three tactile stimuli, and finally assign three different emotions to the three olfactory stimuli. Participants associations did not need to be exclusive (e.g. both \textit{Spiky} and \textit{Round} could be vanilla), and children were not corrected on any of their choices. 
While the section introducing and naming the characters was always presented first and the conclusion was always presented last, the order of the remaining three sections of the story was randomised for each participant. The presentation order of the tactile shapes was also randomised. 
Due to limitations with the legibility of the system's interface, the order of the olfactory stimuli was kept consistent across participants to support reliable transcription. To support the children’s comfort and engagement, the activity occurred in an environment familiar to them, typically a room in their nursery. Before each session, we discussed the procedure with staff to ensure our approach was appropriate, and a member of staff was always present in the room. Their presence likely contributed to the smooth running of the sessions: we did not encounter notable challenges in interacting with the children.

\subsection{Data Collection and Analysis}
Each participant was video and audio recorded using a GoPro
positioned so as to hide the child's face, while still capturing their interaction with the stimulus. The first author manually transcribed 2 hours and 29 minutes of video recordings, organising them
into a spreadsheet, capturing the cross-sensory correspondences given and the reasons for each choice. Not every child explained their reasoning (n = 3 tactile, n = 8 olfactory).
Where children chose a name other than ``Kiki'', ``Bouba'' or``Neither'', these were grouped as ``Other''. 

We calculated descriptive statistics and performed significance testing using Pearson's Chi-square test between variables. Residual analyses were conducted to further explore patterns in the data, following established procedures in previous cross-sensory association studies \cite{dreksler2019critical}. Residuals near zero indicate that the observed frequencies closely match the expected frequencies, suggesting no strong preference. Large positive residuals (usually above 2) suggest that an option was chosen more often than expected by chance, while large negative residuals (usually below -2) indicate it was chosen less often than expected. In the standardised residual tables, we highlight large positive residuals in green and large negative residuals in red. 
Building on prior research into cross-sensory correspondences \cite{likepopcorn, feelingcolours, bluesalt, senseonary,fengItsTouchingUnderstanding2022}, we conducted deductive coding to identify the different association strategies participants used in creating connections between different touch, smell and emotions. \rev{The first author conducted the initial coding, after which the second author reviewed. Any discrepancies between the first and second authors were then discussed between all other authors to reach an agreement.} These association strategies are not mutually exclusive. Children often layered strategies, and in such cases, we coded all of the strategies used.

\section{Findings}
\subsection{Cross-sensory Correspondences}
\begin{figure*}[htbp]
    \begin{subfigure}[b]{0.48\linewidth}
        \includegraphics[width=\linewidth]{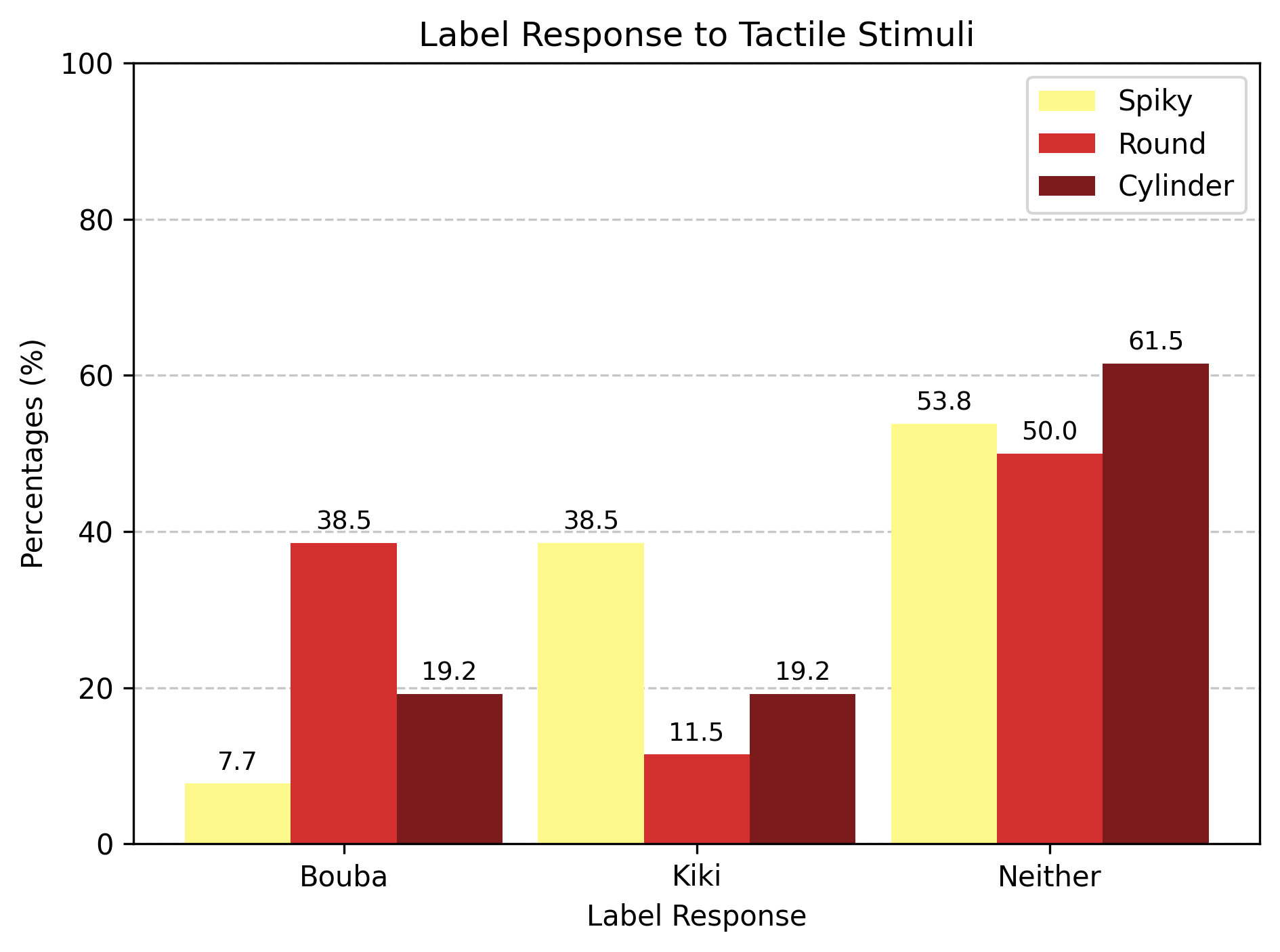}
        \captionsetup{width=.8\linewidth}
        \caption{Names given by children aged 2-4 (N = 26) in response to three different tactile shapes.}
        \label{fig:AllShapeNames25}
        \Description{The bar chart titled ``Label Response to Shape Stimuli'' shows the percentage distribution of different label responses (Bouba, Kiki, Neither, and Other) across three shape categories: Spiky, Round, and Cylinder. The data shows that Round shapes are most frequently associated with the label Bouba, accounting for about 38.5\% of responses, while Spiky shapes are predominantly linked to the label Kiki, also around 38.5\%. Cylinder shapes receive a more varied set of responses, with notable percentages in the Kiki, Neither, and Other categories. Interestingly, the Other label category has the highest percentage for both Round and Cylinder shapes at 50\%, and for Spiky shapes at approximately 46.2\%. These results suggest a strong connection between shape characteristics and the labels assigned, with Round shapes favouring Bouba, Spiky shapes favouring Kiki, and Cylinder shapes showing a less defined pattern.
    \begin{table}[]
    \begin{tabular}{lllll}
     & Spiky & Round & Cylinder & Total \\
    Kiki & 10 & 3 & 5 & 18 \\
    Bouba & 2 & 10 & 5 & 17 \\
    Neither & 14 & 13 & 16 & 42 \\
    Total & 26 & 26 & 26 & 78
    \end{tabular}
    \end{table}}

    \end{subfigure}
     \begin{subfigure}[b]{0.48\linewidth}
        \includegraphics[width=\linewidth]{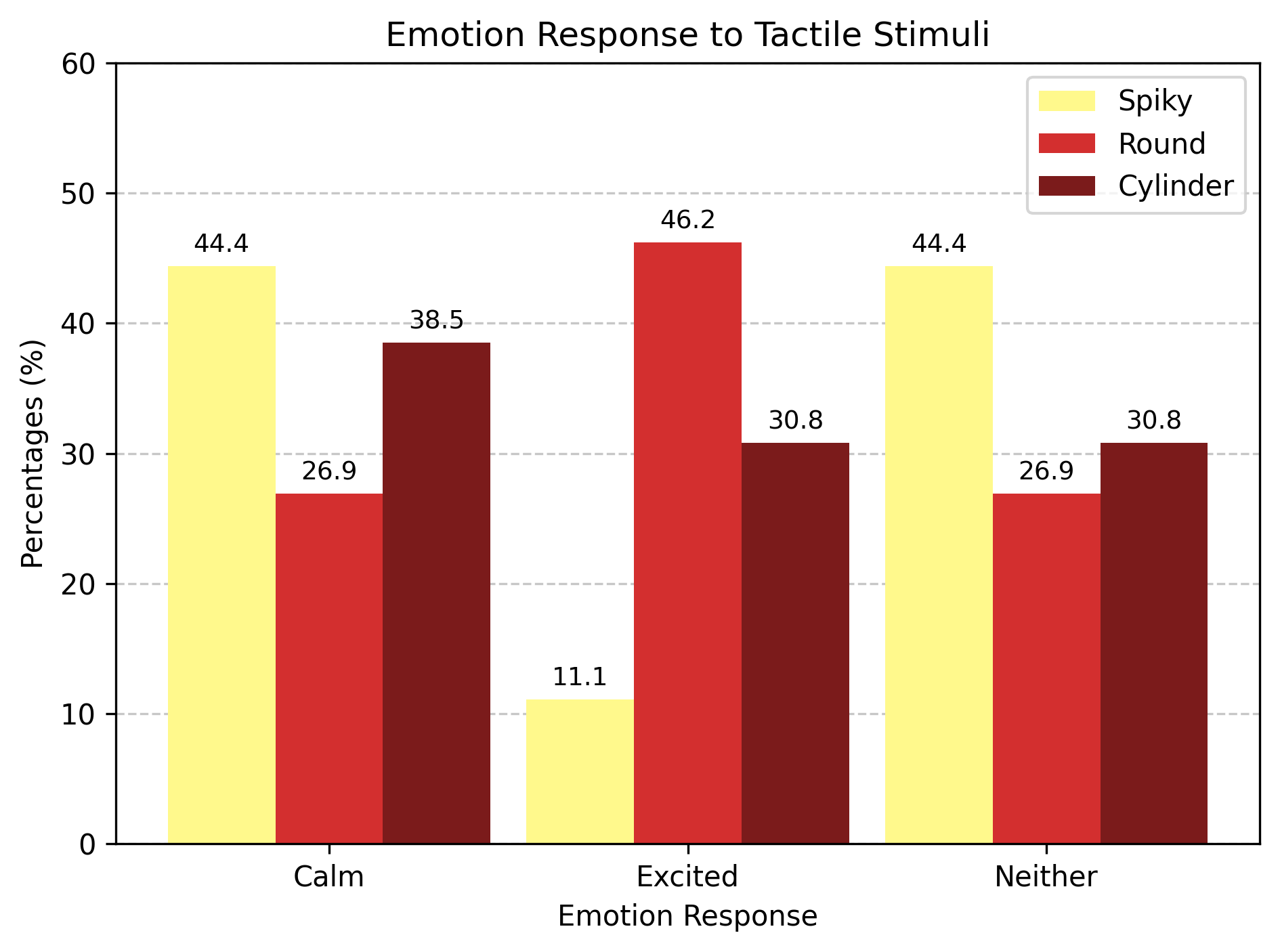}
        \captionsetup{width=.8\linewidth}
        \caption{Emotional responses to three different tactile stimuli with children aged 2-4 (N = 26).}
    \label{fig:AllTactileEmotions25}
    \Description{The bar chart titled``Emotion Response to Shape Stimuli'' shows the percentage distribution of emotional responses (Calm, Excited, and Neither) across three shape categories: Spiky, Round, and Cylinder. The data reveals that Spiky shapes were associated with 44.4\% of participants feeling Calm and an equal 44.4\% feeling Neither, but only 11.1\% feeling Excited. Round shapes are predominantly associated with excitement, receiving the highest percentage of Excited responses at 46.2\%, while Calm and Neither responses are both lower at 26.9\%. Cylinder shapes elicit a more balanced distribution of emotional responses, with 38.5\% Calm, 30.8\% Excited, and 30.8\% Neither. Overall, Spiky shapes tend to evoke calmness or neutrality, Round shapes are linked with excitement, and Cylinder shapes produce a more evenly spread emotional reaction.
    \begin{table}[]
    \begin{tabular}{lllll}
     & Spiky & Round & Cylinder & Total \\
    Calm & 11 & 8 & 11 & 30 \\
    Excited & 3 & 12 & 9 & 24 \\
    Neither & 12 & 6 & 6 & 24 \\
    Column total & 26 & 26 & 26 & 78 \\
    Total & 26 & 26 & 26 & 78
    \end{tabular}
    \end{table}}
    \end{subfigure}
    \hfill
        \begin{subfigure}[b]{0.48\linewidth}
        \includegraphics[width=\linewidth]{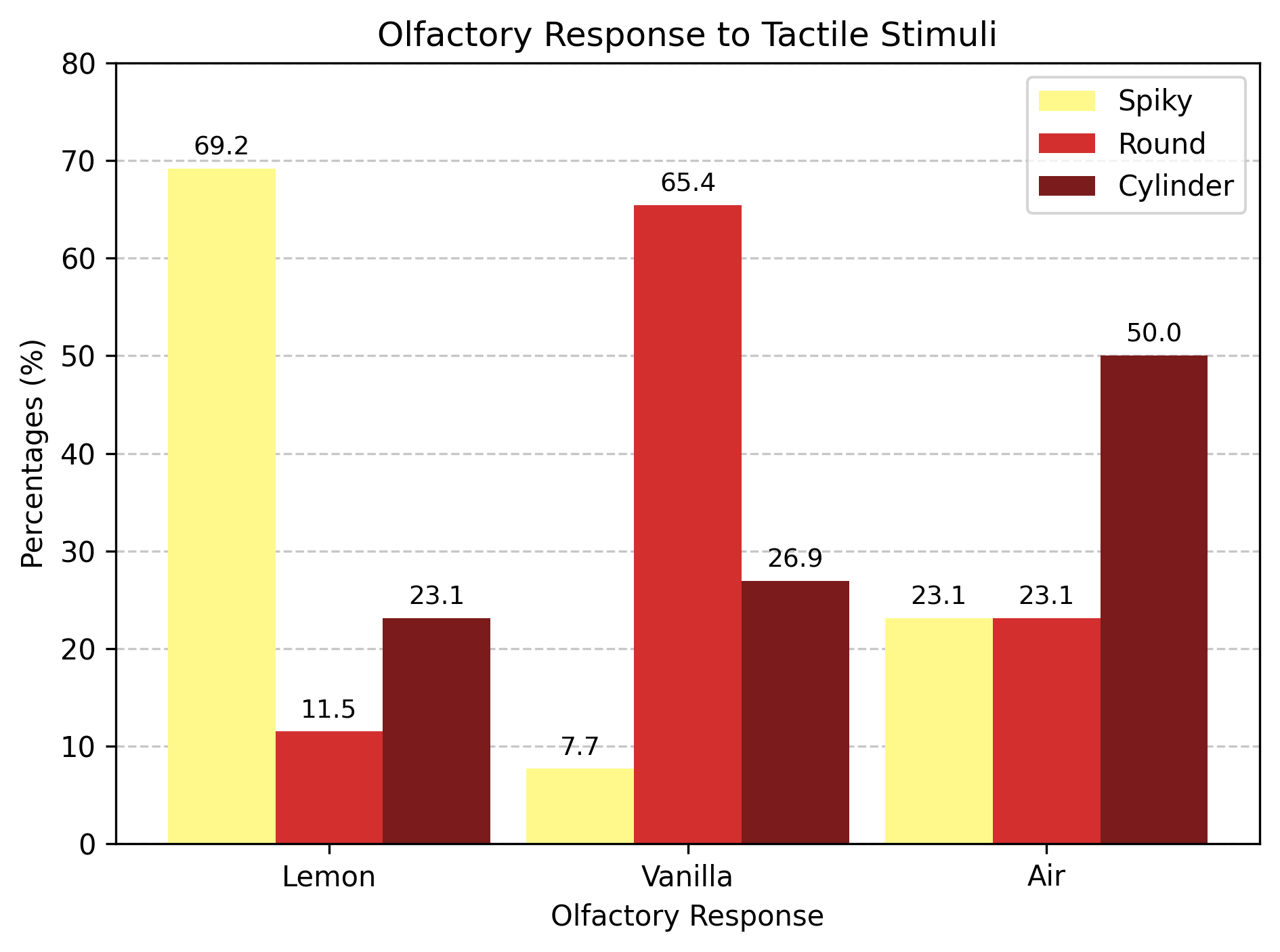}
        \captionsetup{width=.8\linewidth}
        \caption{Distribution of olfactory responses to three tactile stimuli combined in children aged 2-4 (n = 26).}
    \label{fig:AllTactileOlfactory25}
    \Description{The bar chart titled``Olfactory Response to Shape Stimuli'' depicts the percentage of different olfactory responses (Lemon, Vanilla, and Air) associated with three shape categories: Spiky, Round, and Cylinder. The data shows that Spiky shapes are most strongly linked with the Lemon scent, receiving the highest percentage of 69.2\%. Round shapes are predominantly associated with the Vanilla scent, with 65.4\% of responses. Cylinder shapes have a more balanced distribution but are most linked to the Air scent at 50\%. The other associations are lower, with Round shapes having an 11.5\% response for Lemon and 23.1\% for Air, Spiky shapes showing 7.7\% for Vanilla and 23.1\% for Air, and Cylinder shapes having 23.1\% for Lemon and 26.9\% for Vanilla. Overall, the results suggest a strong connection between specific shapes and certain scents, with Spiky linked to Lemon, Round to Vanilla, and Cylinder to Air.
    \begin{table}[]
    \begin{tabular}{lllll}
     & Spiky & Round & Cylinder & Total \\
    Lemon & 18 & 3 & 6 & 27 \\
    Vanilla & 2 & 17 & 7 & 26 \\
    Air & 6 & 6 & 13 & 25 \\
    Column Total & 26 & 26 & 26 & 78 \\
    Total & 26 & 26 & 26 & 78
    \end{tabular}
    \end{table}}

    \end{subfigure}
     \begin{subfigure}[b]{0.48\linewidth}
        \includegraphics[width=\linewidth]{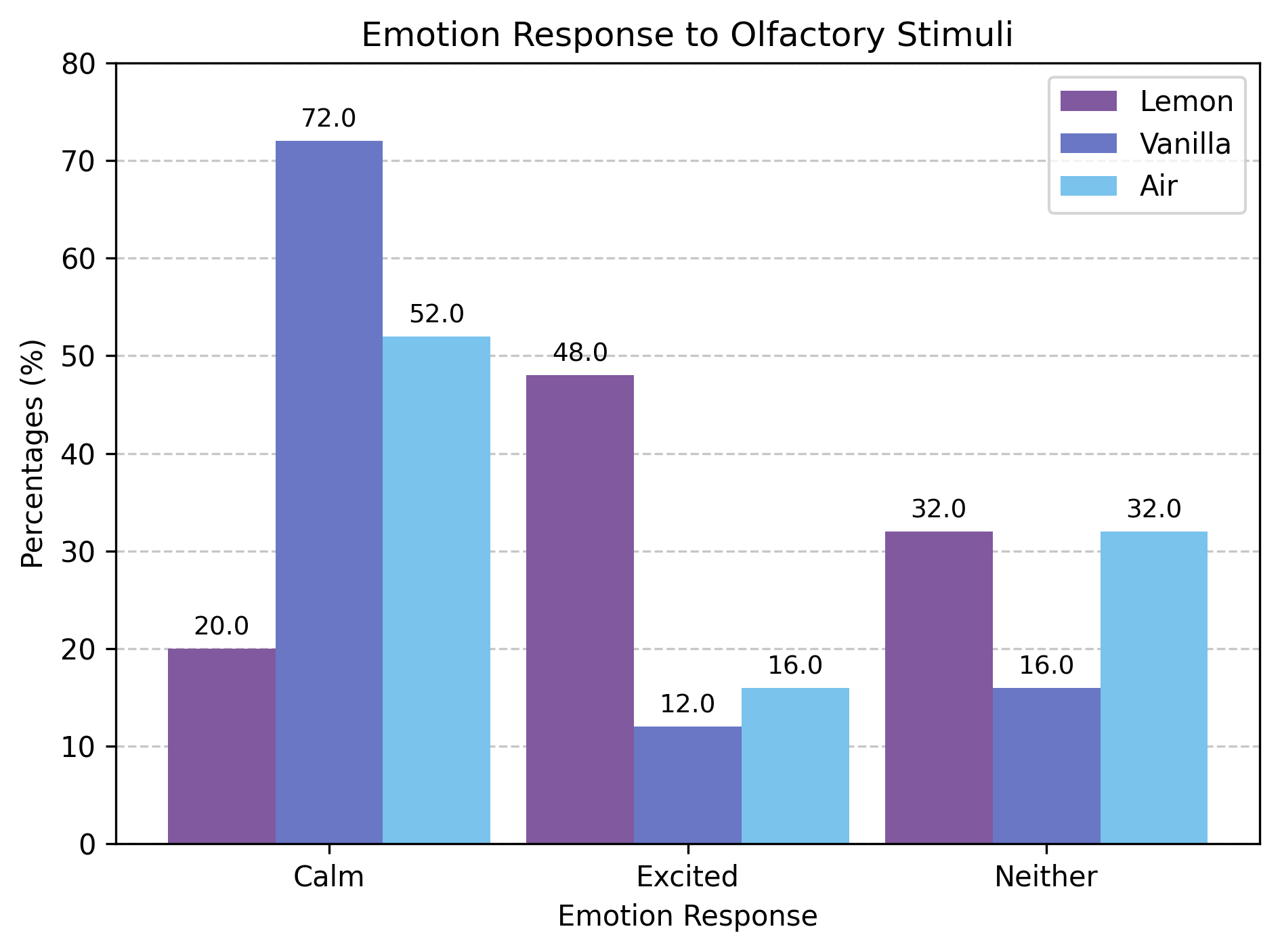}
        \captionsetup{width=.8\linewidth}
        \caption{Distribution of emotional responses to three different scent stimuli in children aged 2-4 (n = 26).}
    \label{fig:AllSmellEmotion25}
    \Description{The bar chart titled``Emotion Response to Olfactory Stimuli'' shows the percentage distribution of emotional responses (Calm, Excited, and Neither) to three different smells: Lemon, Vanilla, and Air. The data indicates that Vanilla scent elicits the highest percentage of Calm responses at 72\%, followed by Air at 52\%, and Lemon at 20\%. In contrast, the Lemon scent provokes the strongest Excited response at 48\%, with Air and Vanilla trailing at 16\% and 12\%, respectively. For the Neither response, Lemon and Air are tied at 32\%, while Vanilla has the lowest percentage at 16\%. Overall, Vanilla tends to induce calmness, Lemon evokes excitement, and Air is associated with calm.
    \begin{table}[]
    \begin{tabular}{lllll}
     & Lemon & Vanilla & Air & Total \\
    Calm & 5 & 18 & 11 & 34 \\
    Excited & 12 & 3 & 5 & 20 \\
    Neither & 8 & 4 & 9 & 21 \\
    Column Total & 25 & 25 & 25 & 75 \\
    Total & 26 & 26 & 26 & 78
    \end{tabular}
    \end{table}}
    \end{subfigure}
    \hfill
    \captionsetup{width=.9\linewidth}
    \hspace{0.35cm}
    \caption{Four bar charts showing the four different associations we researched. We found \textit{statistical significance} in tactile-linguistic, olfaction-emotion and tactile-olfactory. We did \textit{not find statistical significance} in tactile-emotion }
    \label{fig:BarCharts}
\end{figure*}
\subsubsection{Tactile Linguistic Association}

Results are shown in \autoref{fig:AllShapeNames25}, and \autoref{SRTactileLabel}.
The chi-square test of independence indicated a significant association between label and tactile stimuli, $\chi^2(4, N = 78) = 10.42$, $p = 0.034$. 
Calculation of standardised residuals shows that these results were driven by strong associations between ``Kiki'' and \textit{Spiky}, ``Bouba'' and \textit{Round}, and a negative association between ``Bouba'' and \textit{Spiky} (see \autoref{SRTactileLabel}).

\begin{table}[htbp]
\begin{tabular}{|l|l|l|l|}
\hline
\rowcolor[HTML]{B4C6E7} 
\textbf{} & \textbf{Spiky} & \textbf{Round} & \textbf{Cylinder} \\ \hline
``Kiki'' & \cellcolor[HTML]{B3F5B7}2.28 & -1.71 & -0.57 \\ \hline
``Bouba'' & \cellcolor[HTML]{FAA2A2} -2.13 & \cellcolor[HTML]{B3F5B7}2.52 & -0.39 \\ \hline
``Neither'' & -0.16 & -0.64 & 0.80 \\ \hline
\end{tabular}
\Description{The standardised residuals for the labels and tactile stimuli. Spiky shapes show a strong positive association with the label ``Kiki'' (2.28) and a negative association with ``Bouba'' (-2.13). Round shapes show a strong positive association with ``Bouba'' (2.52) and a negative association with ``Kiki'' (-1.71). Cylinder shapes show a moderate positive association with ``Neither'' (1.31) and weak or negative associations with the other labels. The Other category shows only weak associations across all shapes.}
\captionsetup{width=0.9\linewidth}
\caption{Standardised residuals for the association between tactile stimuli (Spiky, Round, Cylinder) and label response (Bouba, Kiki and Neither).}
\label{SRTactileLabel}
\end{table}
\rev{The ``Neither'' option includes instances where children gave a new name as opposed to `Bouba'' or ``Kiki''. This equated to 50\% of the ``Neither'' instances for both \textit{Spiky} and \textit{Round} and 46.2\% for the \textit{Cylinder}.}  Common ``other'' names for the \textit{Spiky} stimulus included ``Spiky''or ``Spike'' (11.5\%) ``Star'' (11.1\%) and ``Spitfire'' (5.6\%).
For the \textit{Round} stimulus: ``Grapes'' (11.8\%), ``Flower'' (11.8\%) ``Bubble'' (5.9\%) and ``Ball'' (5.9\%), and 
for the \textit{Cylinder} stimulus 
``Circle'' (18.8\%), ``Candle'' (6.3\%) and ``Marshmallow''(6.3\%). 



\subsubsection{Tactile Stimuli and Emotion}
A chi-square test of independence indicated no significant association between shape, $\chi^2(4, N = 78) = 8.85$, $p = 0.065$. Associations are plotted in \autoref{fig:AllTactileEmotions25}.
\subsubsection{Olfaction and Emotion}

One participant did not answer this set of questions, so for this analysis, n = 25. 
A chi-square test of independence indicated a significant association between olfactory stimuli, $\chi^2(4, N = 75) = 16.17$, $p < 0.005$.
Calculation of standardised residuals shows that these results were driven by strong positive associations between vanilla and calm,  lemon and excited, and negative association between lemon and calm, and vanilla and excited (see \autoref{SROlfactoryEmotion}).
Associations are plotted in \autoref{fig:AllSmellEmotion25}.

\begin{table}[htbp]
\begin{tabular}{|l|l|l|l|}
\hline
\rowcolor[HTML]{B4C6E7} 
\textbf{} & \textbf{Lemon} & \textbf{Vanilla} & \textbf{Air} \\ \hline
Calm & \cellcolor[HTML]{FAA2A2}-3.12 & \cellcolor[HTML]{B3F5B7}3.28 & 1.10 \\ \hline
Excited & \cellcolor[HTML]{B3F5B7}2.95 & \cellcolor[HTML]{FAA2A2}-2.03 & -1.78 \\ \hline
Neither & 0.55 & -1.64 & 0.33 \\ \hline
\end{tabular}
\captionsetup{width=0.9\linewidth}
\caption{Standardised residuals for the association between olfactory stimuli (Lemon, Vanilla, Air) and emotional response (Calm, Excited, Neither).}
\label{SROlfactoryEmotion}
\Description{ The standardised residuals for the olfactory and emotions. For the lemon stimulus, the standardised residuals show a strong negative association with calm responses (-3.12), a strong positive association with excited responses (2.95), and a weak positive association with neither responses (0.55). For the vanilla stimulus, the standardised residuals indicate a strong positive association with calm responses (3.28), a strong negative association with excited responses (-2.03), and a weak negative association with neither responses (-1.64). For the air stimulus, the standardised residuals show a slight positive association with calm responses (1.10), a negative association with excited responses (-1.78), and a weak positive association with neither responses (0.33).}
\end{table}

\subsubsection{Tactile Stimuli and Olfactory}
A chi-square test of independence indicated a significant association between tactile stimuli (Spiky, Round, Cylinder) and olfactory Stimuli (Lemon, Vanilla, Air), $\chi^2(4, N = 78) = 31.38$, $p < 0.001$. 
Calculation of standardised residuals shows that these results were driven by very strong positive associations between lemon and spiky, round and vanilla, and a strong association between cylinder and air, as well as strong negative associations between vanilla and spiky, and round and lemon
(see \autoref{SROlfactoryEmotion}).
Associations between tactile and olfactory stimuli in children are plotted in Figure \ref{fig:AllTactileOlfactory25}. 


\begin{table}[htbp]
\begin{tabular}{|l|l|l|l|}
\hline
\rowcolor[HTML]{B4C6E7} 
\textbf{} & \textbf{Spiky} & \textbf{Round} & \textbf{Cylinder} \\ \hline
Lemon & \cellcolor[HTML]{B3F5B7}4.54 & \cellcolor[HTML]{FAA2A2}-3.03 & -1.51 \\ \hline
Vanilla & \cellcolor[HTML]{FAA2A2}-3.40 & \cellcolor[HTML]{B3F5B7}4.25 & -0.85 \\ \hline
Air & -1.20 & -1.20 & \cellcolor[HTML]{B3F5B7}2.40 \\ \hline
\end{tabular}
\captionsetup{width=0.9\linewidth}
\caption{Standardised residuals for the association between tactile stimuli  (\textit{Spiky}, \textit{Round}, \textit{Cylinder}) and olfactory response (lemon, vanilla, air). }
\label{RSTactileOlfactorty}
\Description{This table shows the strength of association between olfactory responses and 3D shape types. Spiky shapes are strongly associated with Lemon (4.54) and negatively associated with Vanilla (-3.40). Round shapes show a strong positive association with Vanilla (4.25) and a negative association with Lemon (-3.03). Cylinder shapes show a moderate positive association with Air (2.40) and weaker negative associations with the other scents.}
\end{table}

\subsection{Cross-sensory Association Strategies}

\begin{figure*}[htbp]
\captionsetup{width=0.9\linewidth}
    \centering
    \includegraphics[width=\linewidth]{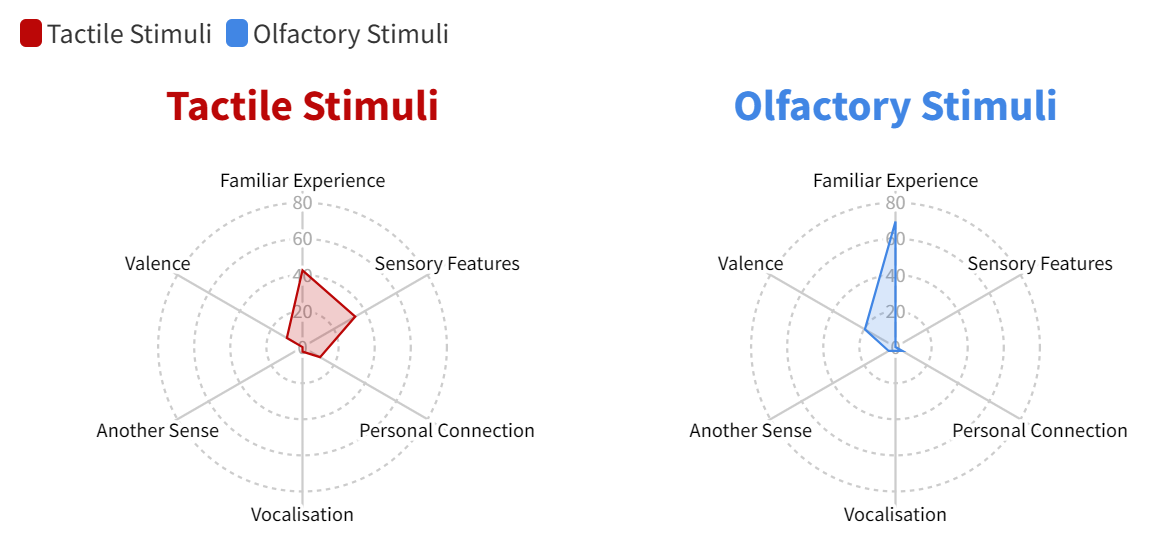}
    \caption{Radar charts showing the percentage of association strategies used to explain correspondences for tactile (left) and olfactory (right) stimuli.}
    \Description{For tactile stimuli, participants most often relied on familiar experiences (42.5\%) and sensory features (33.8\%) when explaining their correspondences, with smaller proportions using personal connections (11.3\%), valence (10\%), or vocalisations (2.5\%), and none referring to another sense. In contrast, olfactory stimuli were explained predominantly through familiar experiences (69.6\%), followed by valence (19.6\%), with much lower use of personal connections (4.4\%), another sense (4.4\%), or vocalisations (2.2\%), and no references to sensory features.
    \begin{table}[]
\begin{tabular}{lllllll}
Association   Strategy & Familiar   Experience & Sensory Features & Personal   Connection & Vocalisation & Another Sense & Valence \\
Tactile Stimuli & 42.50\% & 33.75\% & 11.25\% & 2.50\% & 0.00\% & 10.00\% \\
Olfactory Stimuli & 69.57\% & 0.00\% & 4.35\% & 2.17\% & 4.35\% & 19.57\%
\end{tabular}
\end{table}}
    \label{fig:placeholder}
\end{figure*}
For the tactile stimuli, participants gave a total of 80 reasons electing the correspondences they did (over both name and emotion). Overall, the most frequently employed association strategies were \textit{Familiar Experience} (42.50\%) and \textit{Sensory Features} (33.75\%). Examination of strategy use by specific tactile stimuli reveals notable differences. For the \textit{Spiky} stimulus, children predominantly relied on \textit{Familiar Experience}, referencing recognisable items like ``star'' and ``a diamond'', and \textit{Sensory Features}, using words like ``sharp'', or ``hard'' (both 40.00\%).

For the \textit{Round} stimulus, children relied on \textit{Familiar Experience} (48.28\%), referencing items such as ``grapes'' and ``bouncy balls'', but showed increased instances of \textit{Personal Connection} (13.79\%) and \textit{Valence} (13.79\%) responses, using phrases such as ``I love it''. With the \textit{Cylinder} stimulus, children mainly relied on \textit{Sensory Features} (42.86\%), referencing ``water'' and ``Lego sets'', similar to that observed for the \textit{Spiky} stimulus, but fewer references reflecting \textit{Personal Connection}. Participants also explained their association in terms of \textit{Sensory features} (33.75\%), using words like ``hard'' and ``smooth''. Interestingly, the word ``friend'' (n = 9) emerged primarily in relation to emotional associations; for example, one participant (P4) explained that the \textit{Round} stimulus elicited excitement because ``they have friends'' whereas the Spiky stimulus elicited calm because ``they have no friends''.

For the olfactory stimuli, participants gave a total of 46 reasons for selecting correspondences they did (over both emotion and tactile). Across all three scents, the most frequently employed strategies were \textit{Familiar Experience} (69.57\%) and \textit{Valence} (19.57\%). Examination of strategy use by specific olfactory stimuli reveals notable differences. For the lemon stimuli, children predominantly used \textit{Familiar Experience} (80.00\%), referencing familiar items like ``strawberries'', ``flowers'' and ``Play-Doh''. The only other strategies utilised were\textit{Valence} (5.00\%), where phrases such as ``I don't like them'' were used, \textit{Personal Connection} (5.00\%), where one participant (P15) explained that her ``daddy makes Play-Doh that smells like this'' which was why she associated lemon with excited, and \textit{Grasping for Another Sense} where one child said both stimuli``taste like fruit''. For the vanilla stimuli, all strategies were used other than \textit{Sensory Features}, with \textit{Familiar Experience} being used predominantly (57.14\%), referencing items such as ``lollipop'' and ``toothpaste''. \textit{Valence} was also relied on (21.43\%) using phrases such as ``I love it'', and ``I like it''. For the air stimulus, only two association strategies were relied on when explaining associations: \textit{Familiar Experience} (66.67\%), with references to ``salt'' and ``air'' and \textit{Valence} (33.33\%), using phrases such as ``I love it'' and ``I don't like''. Notably, the term ``friend'' appeared multiple times (n = 6), paralleling findings with tactile stimuli, and was used to frame emotional associations. However, there were no instances of the \textit{Sensory features} strategy being used, which differs strongly from the tactile stimuli.

\section{Discussion}

\subsection{Developmental Patterns in Cross-Sensory Correspondences}

Compared to previous findings in older age groups, our results reveal both continuities and divergences in how preschoolers formed cross-sensory correspondences, likely indicative of the effect of early development on sensory integration.
For tactile-linguistic associations, we observed a clear ``Bouba–Kiki'' effect: children reliably matched the \textit{Spiky} stimulus with ``Kiki’’ and the \textit{Round} stimulus with ``Bouba’’, consistent with prior findings on visual–auditory correspondences in 2D shapes \cite{maurer2006shape}. This suggests both that our stimuli are comparable to prior cross-sensory work and that some correspondences may emerge early and remain stable across modalities and developmental stages. \rev{Overall, the most common response was for children to select ``neither'' then propose another name. We found that when a participant gave a new name to one shape, they would give a new name to all shapes (n = 13). Interestingly, the names suggested still conformed to the phonetic patterns associated with the bouba–kiki effect. For the angular stimulus, \textit{Spiky}, children generated labels such as ``Spiky'', ``Spike'', ``Star'', and``Spitfire'', all of which include consonants such as /k/, /t/ and /p/ \cite{mccormick2015sound} that are typically associated with angularity. For the \textit{Round} stimulus, children produced names like ``Flower'', ``Bubble'', and ``Ball'', which feature consonants such as /l/, /b/ and /u/ \cite{mccormick2015sound} commonly linked to bouba-type roundedness. The names given also mapped onto the association strategies in our coding scheme, in particular \textit{Familiar Experience} and \textit{Sensory Features}.}

In contrast, tactile-emotion mappings showed no statistical significance and, in fact, tended to the opposite association observed in older children and adults \cite{likepopcorn, fengItsTouchingUnderstanding2022}. Whereas adults and older children typically associate spikiness with high arousal and roundness with calm, preschoolers sometimes described the \textit{Spiky} shape as calm and the \textit{Round} shape as excited. This may reflect the less developed stage of socio-emotional and linguistic expression in this age group \cite{KestenbaumGelman1995, bender2011young, Knothe2023,Grosse2021}.
Children often explained their reasons for these associations in anthropomorphic terms: \textit{Spiky} had ``no friends’’ whereas \textit{Round} ``had friends’’, suggesting that they grounded emotional reasoning in social terms. This may have been encouraged by our narrative framing of the stimuli as characters on an adventure, but also reflects young children’s 
tendency to personify objects and explain via social reasoning \cite{goldman2024children, bluesalt}. 

For olfactory-emotion associations, children showed clear and statistically significant correspondences, linking vanilla with calm and lemon with excited. These findings align closely with prior research on older children and adults \cite{likepopcorn}, where sweeter, pleasant odours elicit low-arousal emotions and sharper, citrus scents elicit higher-arousal states. The stability of these correspondences across age groups suggests that smell may provide a robust and potentially universal channel for mapping sensory input to emotion. 
Finally, tactile-olfactory mappings also showed significance, with \textit{Spiky} paired with lemon and \textit{Round} with vanilla. This mirrors findings in adults \cite{spencesmell}.
However, many preschoolers struggled to articulate explicit reasoning for these choices, indicating that while such correspondences may already operate intuitively, they are not yet consciously accessible or easily verbalised. 

\rev{In developing our study, we hoped that the pre-schoolers would provide verbal justifications for their choices. However, during the study, we found that children gave no verbal explanation or only a brief explanation. This may be due to cross-sensory correspondences operating at an intuitive level \cite{CrossmodalCorrespondeces, spence2023explaining} and emotions being difficult for young children to verbalise \cite{widen2008young}. Children may have relied more on implicit associations rather than explicit reasoning. 

We also expected toddlers to have similar correspondences as adults. Our data suggested this was the case in all tasks other than the tactile-emotion task: while adults associate angular shapes with excitement, and rounded shapes with calmness, the children in our study showed the opposite association.} Taken together, these findings suggest that some correspondences (e.g., tactile-linguistic, olfactory-emotion, tactile-olfactory) may appear early and remain stable across development, while others (e.g., tactile-emotion) may still be in development and grounded in social imagination at this age, only later aligning with adult patterns. This underscores the need to situate cross-sensory research in early childhood not only within perceptual mechanisms but also within the developmental context of social and narrative reasoning.

\subsection{How Preschoolers Reason About Cross-Sensory Correspondences}

Our analysis of association strategies provides insight into how preschoolers reason about and articulate the cross-sensory correspondences they experience, but does not offer a definitive categorisation. Again, these results indicate both continuities and divergences from results in older children and adults.

Across both tactile and olfactory tasks, the \textit{Familiar Experience} strategy was most common. Children frequently grounded their associations in everyday objects, e.g. calling the \textit{Round} shape a ``grape’’. This echoes prior findings with older children \cite{senseonary, bluesalt}, but stands in contrast to adults, who tend to draw on more abstract or culturally learned metaphors \cite{piaget1952origins}. This suggests that at preschool age, sensory reasoning is still tightly coupled with concrete, lived experiences. 

For olfactory stimuli, children also explained their associations in terms of \textit{Valence} (e.g., whether they liked or disliked the stimuli). When explaining their associations with Vanilla, children often invoked positive valences. For example, when asked why vanilla felt calm, one child responded ``I love it’’. Similarly, lemon was often associated with negative valences. This is consistent with evidence for the influence of odours on mood and affective cognition \cite{herz2002influences}. Notably, 
no children explained their olfactory associations in terms of \textit{Sensory Features}.
One possible explanation is linguistic: the English language offers a limited vocabulary for describing smells compared to other senses \cite{lostintranslation, digonnet2018linguistic}, making valence-based reasoning more accessible than feature-based description.

In contrast, reasons for associations with tactile stimuli were often described in terms of  \textit{Sensory Features}, with children describing shapes as ``spiky’’ or ``round’’. Whereas older children in prior work used sensory descriptions relatively rarely \cite{bluesalt}, our preschool-aged participants used them in over a third of cases. This greater reliance on direct sensory features may reflect developmental differences in exploratory behaviour. At this age, tactile exploration is a primary mode of learning, with children often relying on immediate properties such as texture and shape to construct meaning \cite{Gillioz2024, multiEdu, shams2008benefits}. 

The above patterns suggest that preschoolers’ cross-sensory association strategies are grounded in immediate experience (familiar objects, perceptual features, and affective responses). Unlike adults, who may rely on abstract conventions, young children root their reasoning in what they can touch, see, or feel emotionally in the specific moment of interaction \cite{woodward2001infants}. 
While this would require confirmation in future work, this suggests a possible trajectory in the developmental trajectory of cross-sensory reasoning:
starting in concrete and embodied experiences, and gradually expanding to abstract, then metaphorical, and culturally shared associations as we age.

\subsection{Methodological Lessons for Studying Cross-Sensory Correspondences with Preschoolers}

While prior work with adults has examined a wide range of sensory mappings \cite{steer2024squishy, masie, feelingcolours, bubbletea}, these methods are often unsuitable for younger children, who are at a critical stage of sensory integration. To address this gap, we developed and trialled a replicable method tailored to 2-4-year-olds.
As such our work offers methodological insights for future work seeking to study sensory and affective experiences
with younger children. 
\rev{During the study, we did not experience any significant issues. Having intentionally simplified the task and embedded it within the story, we found that pre-schoolers (n = 25) were able to complete all the tasks. In the one case where a child did not complete all of the tasks this was due to the researcher forgetting to ask one of the questions (the participant was excluded). We believe the success of the protocol was also improved by the familiarity of the environments, and the presence of staff the children knew and felt comfortable with.}

\rev{A key aspect of our approach was the embedding of sensory tasks within a story-based format --- intended to support the children's engagement and help them articulate the correspondences they perceived. 
While we did not formally evaluate the impact of this choice on engagement and compliance, the approach appeared effective for a number of reasons. 
No children wanted to leave during an experiment, and we observed that they remained focused throughout the session, in ways that children often do not in experimental studies. On leaving the activity, many of the children reported enjoying the activity. 
This was likely also helped by our decision to limit the number of stimuli to reduce cognitive and memory demands.
More specifically to our particular study, the narrative 
approach also allowed us to incorporate natural breaks from olfactory stimulation, while keeping the children occupied and engaged. 
Our experience suggests that our storytelling framing and other design choices may offer practical and effective approaches for future studies with children. However, it will be necessary for future work to evaluate the benefits of these design choices more formally.}

At the same time, our method highlighted inclusivity challenges. Because our experimental process required children to produce verbal responses, non-verbal preschoolers were excluded. Future research could extend accessibility by introducing embodied or tangible response options, such as arranging objects or pointing, which would allow associations to be captured without relying exclusively on verbal explanation. While these approaches may make reasoning harder to interpret, they would broaden participation to children with diverse communicative profiles. Story-based designs could also provide a flexible scaffold for these alternative modes of response, ensuring that children of varying abilities can take part.

We thus suggest that story-driven, simplified, and sensory methods may offer a practical and replicable approach for studying cross-sensory correspondences with very young children. We see potential for extending this approach to other sensory domains (e.g. taste, audition) and to more inclusive populations, thus broadening the methodological foundations for sensory HCI research with preschoolers.

\subsection{Design Insights for Cross-Sensory Technology}

Our findings carry several implications for the design of child-centred cross-sensory and multisensory technologies. First, designers should carefully consider the age of the user. For preschoolers, interactions can be made effective by grounding them in \textit{Familiar Experiences} and in sensory qualities that are clear, exaggerated, and easy to perceive. Consistent with recent work on learning technologies \cite{farming}, methodological effectiveness of storytelling for empirical studies can also be extended into a principle for technology design --- to engage children and support their ability to articulate and reason about experiences.

Designers must be attentive to the meanings children attach to sensory combinations. For example, in our study, lemon was consistently associated with excitement, while vanilla was linked with calm. 
Where adults may have formed more complex and situational associations with scents (e.g. a relaxing lemon and ginger tea) using lemon, or more broadly citrus scents with preschoolers in a calm context, or to signal a calm character may therefore create confusion. Such findings highlight the importance of aligning cross-sensory cues with children’s expected correspondences rather than adult expectations, particularly when designing for affective communication.
Additionally, there is an opportunity for design to help children engage in a wider, richer range of sensory expression, beyond their common recourse to \textit{Familiar Experiences} in our results. Technologies could, for instance, provide prompts or tools that expand children’s ability to describe smells beyond everyday references, supporting the development of a broader vocabulary for olfaction \cite{MASON201948}. This could also enrich how children communicate their experiences, opening new possibilities for creative and educational uses of smell in interaction.

\rev{Building on these points, the stable mappings between emotion–olfaction, olfaction–tactile, and tactile–linguistic could be leveraged to create assistive tools that support children, particularly those with limited verbal abilities, communicate how they feel by choosing scents that correspond with their emotional states. These correspondences also have clear value in storytelling and educational media, where aligning shapes and smells, smells and emotions, and shapes and names can scaffold children’s developing emotional literacy and support richer sensory engagement.}

Finally, our results suggest that cross-sensory correspondences could be leveraged to support emotional communication, e.g. for children who are non-verbal. Preschoolers in our study successfully mapped emotions onto olfactory stimuli, indicating that cross-sensory cues may provide alternative pathways for children to express internal states that are otherwise difficult to articulate, \rev{by using smell as a pathway}. This opens promising avenues for the design of inclusive technologies that extend beyond entertainment or education to support emotional well-being and social connection.

\subsection{Limitations and Future Work}

While our study provides new insights into cross-sensory correspondences in preschoolers, several limitations should be acknowledged. First, some children struggled to provide explicit reasons for their tactile and olfactory associations \rev{(n = 3 tactile, n = 8 olfactory)}. Although this reflects developmental constraints in verbal reasoning, it also highlights the need for future work to explore more diverse ways of capturing children’s expression. For instance, embodied or tangible response options, such as arranging objects, pointing, or demonstrating through play, could support participation by children who are non-verbal or less comfortable with verbal explanations, while still revealing the associations they form.

While the number of stimuli in our study was intentionally limited to reduce cognitive load and respect sample size constraints, this constrained the range of possible correspondences. Expanding the variety of tactile and olfactory stimuli in future studies could provide a richer understanding of how preschool-aged children map cross-sensory experiences. Longitudinal designs could further explore how these correspondences and association strategies evolve with age, tracing developmental trajectories as children move from intuitive, experience-based associations to more abstract and culturally shaped associations. Furthermore, our analysis of association strategies captured a way to compare children’s reasoning, but categorising layered or overlapping responses simplified their expressions. Future research should investigate how children combine strategies to capture more dynamic and complex ways of making sense of sensory correspondences. 

Finally, while we focused on tactile and olfactory correspondences, other sensory domains also remain underexplored in preschool-aged children, particularly taste and audition. There is, therefore, room to extend story-based and developmentally sensitive methods to these domains, which could open new avenues for sensory CCI research.

\section{Conclusion}
We presented a study with 26 preschoolers investigating the cross-sensory correspondences they formed between touch, smell, and emotion. We found significant associations across tactile–linguistic, tactile–olfactory, and olfactory–emotion, and we examined the strategies preschoolers used to explain these mappings. Our contributions include design guidelines for developing sensory interfaces that align with how preschoolers integrate sensory input, as well as a replicable method for exploring cross-sensory correspondences in early childhood.

\begin{acks}
This work is funded by an ERC Consolidator fellowship/UKRI grant no.: EP/Y023676/1. We would like to thank all participants, parents and teachers that took part and supported this study. 
\end{acks}
\bibliographystyle{ACM-Reference-Format}
\bibliography{biblography}

\appendix

\end{document}